\documentclass[a4paper,12pt]{article}
\usepackage[latin1]{inputenc}
\usepackage{amssymb}

\title{{\bf Topology and ambiguity \\in $\omega$-context free languages } }
\author{Olivier Finkel\\{\it Equipe de Logique Math\'ematique }  
 \\ U.F.R. de Math\'ematiques, Universit\'e Paris 7 \\ {\it 2 Place Jussieu 75251 Paris
 cedex 05, France.}
\\E Mail: finkel@logique.jussieu.fr
\vspace{05mm}
\\ Pierre Simonnet \\ {\it UMR CNRS 6134} \\
 Facult\'e des Sciences, Universit\'e de Corse\\
 {\it Quartier Grossetti BP52 20250, Corte, France } 
\\ E Mail: 
simonnet@univ-corse.fr.} 
\date{}
\begin{document}

\newtheorem{The}{Theorem}[section]
\newtheorem{Pro}[The]{Proposition}
\newtheorem{Deff}[The]{Definition}
\newtheorem{Lem}[The]{Lemma}
\newtheorem{Rem}[The]{Remark}
\newtheorem{Exa}[The]{Example}
\newtheorem{Cor}[The]{Corollary}
\newtheorem{Fact}[The]{Fact}

\newcommand{\fa}{\forall}
\newcommand{\Ga}{\Gamma}
\newcommand{\Gas}{\Gamma^\star}
\newcommand{\Si}{\Sigma}
\newcommand{\Sis}{\Sigma^\star}
\newcommand{\Sio}{\Sigma^\omega}
\newcommand{\Gao}{\Gamma^\omega}
\newcommand{\ra}{\rightarrow}
\newcommand{\hs}{\hspace{12mm}

\noi}
\newcommand{\lra}{\leftrightarrow}
\newcommand{\la}{language}
\newcommand{\ite}{\item}
\newcommand{\Lp}{L(\varphi)}
\newcommand{\abs}{\{a, b\}^\star}
\newcommand{\abcs}{\{a, b, c \}^\star}
\newcommand{\ol}{ $\omega$-language}
\newcommand{\orl}{ $\omega$-regular language}
\newcommand{\om}{\omega}
\newcommand{\nl}{\newline}
\newcommand{\noi}{\noindent}
\newcommand{\tla}{\twoheadleftarrow}
\newcommand{\de}{deterministic }
\newcommand{\proo}{\noi {\bf Proof.} }
\newcommand {\ep}{\hfill $\square$}

\maketitle

\begin{abstract} 
\noi We study the links between the topological complexity of an  
$\om$-context free language  and its degree of ambiguity. 
In particular, using known facts from classical 
descriptive set theory, we prove that non Borel $\om$-context free languages 
 which are recognized by B\"uchi 
pushdown automata have a maximum degree of ambiguity.  
This result implies that degrees of ambiguity are really  not preserved by the operation 
$W \ra W^\om$, defined over finitary context free languages. 
We prove also that taking the adherence or the $\delta$-limit of a finitary language 
preserves neither ambiguity nor inherent ambiguity. 
 On the other side we show that methods used in the study of 
$\om$-context free languages 
can also be applied to study the notion of ambiguity in  
infinitary rational relations accepted by B\"uchi 2-tape automata 
and we get first results in that direction. 
\end{abstract}

\noi {\small {\bf Keywords:} context free languages; infinite words; 
 infinitary rational relations; ambiguity; 
degrees of ambiguity; topological properties; borel hierarchy; analytic sets.}

\hs {\small {\bf AMS Subject Classification:} 68Q45; 03D05; 03D55; 03E15. 

\section{Introduction}
$\om$-context free languages ($\om$-CFL) form the class $CFL_\om$ 
of \ol s accepted by pushdown automata with a B\"uchi or Muller acceptance 
condition. 
They were firstly studied by Cohen and Gold, Linna, Boasson, Nivat, \cite{cg} \cite{lin76} 
\cite{bn} \cite{ni77}, see  Staiger's paper for a survey of these works \cite{sta}. 
A way to study the richness of the class $CFL_\om$ is to consider the 
topological 
complexity of $\om$-context free languages when the set $\Sio$ 
of infinite 
words over the alphabet $\Si$ is equipped with the usual Cantor topology. 
It is well known  that all $\om$-CFL as well as all \ol s accepted by 
Turing machines with a B\"uchi or a Muller acceptance condition  are analytic sets. 
$\om$-CFL accepted by \de B\"uchi pushdown automata are ${\bf \Pi^0_2 }$-sets, while 
$\om$-CFL accepted by \de Muller pushdown automata are boolean 
combinations of ${\bf \Pi^0_2 }$-sets. 
It was recently proved that the class $CFL_\om$ exhausts the finite 
ranks of the Borel hierarchy, \cite{fina}, that there exists some 
$\om$-CFL which are Borel sets of infinite rank, \cite{finc}, or even 
analytic but non Borel sets, \cite{finb}. 

\hs 
Using known facts  from Descriptive Set Theory, 
we prove here that non Borel $\om$-CFL have a 
maximum degree of ambiguity: 
if $L(\mathcal{A})$ is a non Borel $\om$-CFL which is accepted by a 
B\"uchi pushdown automaton  (BPDA) $\mathcal{A}$ then 
there exist $2^{\aleph_0}$  $\om$-words $\alpha$ such that $\mathcal{A}$  
has $2^{\aleph_0}$  accepting runs reading $\alpha$, where $2^{\aleph_0}$ is the cardinal 
of the continuum. 

\hs
 The above  result of the second author led the first author to the   investigation 
of the notion of ambiguity and of degrees of ambiguity in $\om$-context free languages, 
\cite{ambcf}. There exist some non ambiguous $\om$-CFL of every finite Borel rank, 
but all known examples of $\om$-CFL which are Borel sets of infinite rank 
are accepted by ambiguous 
BPDA. Thus one can make the hypothesis that there are some links 
between the topological complexity and the degree of 
ambiguity for $\om$-CFL and such connections were firstly studied in \cite{ambcf}.  
\nl The operations $W\ra Adh(W)$ and $W\ra W^\delta$, where $Adh(W)$ is the
adherence
of the finitary language $W \subseteq \Sis$ and $W^\delta$ is the 
$\delta$-limit of $W$, appear in the
characterization of ${\bf \Pi^0_1 }$ (i.e. closed)-subsets 
and ${\bf \Pi^0_2 }$-subsets of $\Si^\om$, for an alphabet $\Si$, \cite{sta}.
 Moreover it turned out 
that the first one is  useful in the study of topological properties of 
$\om$-context free languages of a given degree of ambiguity \cite{ambcf}.
We show that each of these operations preserves
neither unambiguity nor inherent ambiguity from finitary to $\omega$-context
free languages. 
We deduce also from the above results  
that neither unambiguity nor inherent ambiguity is preserved by 
the operation $W\ra W^\om$. This important operation is defined over finitary languages and 
is involved  in the characterization of the class  of \orl s (respectively,
of $\om$-context
free languages) as the $\om$-Kleene closure of the class of regular (respectively, context
free) languages  \cite{tho} \cite{pp} \cite{sta} \cite{staop}.

\hs
 On the other side we prove that the same theorems of classical descriptive set theory 
can also be applied in the case of infinitary rational relations accepted by 
$2$-tape B\"uchi automata. 
The topological complexity of infinitary rational relations 
has been studied by the first author who showed in \cite{relrat} that there exist
some infinitary rational relations which are not Borel.  Moreover some undecidability 
properties have been established in \cite{rel-dec}. 
We then prove some first results about ambiguity in infinitary rational relations. 

\hs 
 The paper is organized as follows. In section 2, we recall 
definitions and results about $\om$-CFL and ambiguity. In section 3, Borel and analytic sets 
are defined. In section 4, we study links between topology and ambiguity in  $\om$-CFL. 
In section 5, we show some results about 
infinitary rational relations. 

\section{$\om$-context free \la s}

We assume the reader to be familiar with the theory of formal \la s and 
of \orl s,  \cite{bers} \cite{tho} \cite{sta} \cite{pp}. 
We shall use usual notations of formal language theory. 
When $\Si$ is a finite alphabet, a non-empty finite word over $\Si$ is any 
sequence $x=a_1\ldots a_k$, where $a_i\in\Sigma$ 
for $i=1,\ldots ,k$, and $k$ is an integer $\geq 1$. The length 
of $x$ is $k$, denoted by $|x|$ . 
We write $x(i)=a_i$ and $x[i]=x(1)\ldots x(i)$ for $i\leq k$. We write also 
$x[0]=\lambda$, where $\lambda$ is the empty word,  which  has no letter; its length is 
$|\lambda|=0$.   
$\Sis$ is the set of finite words over $\Sigma$, 
and $\Si^+$ is the set of finite non-empty words 
over $\Sigma$. The mirror image of a finite word $u$ will be denoted by $u^R$. 

\hs 
The first infinite ordinal is $\om$. 
An $\om$-word over $\Si$ is an $\om$ -sequence $a_1 \ldots a_n \ldots$, 
where $\fa i\geq 1$~
$a_i \in\Sigma$. 
The set of $\om$-words over the alphabet $\Si$ is denoted by $\Si^\om$. 
An $\om$-language over an alphabet $\Sigma$ is a subset of $\Si^\om$. 
For $V\subseteq \Sis$, the $\om$-power of $V$ is the \ol~ 
$V^\om = \{ \sigma =u_1\ldots u_n \ldots \in \Si^\om \mid 
 \fa i\geq 1 ~u_i\in V - \{\lambda\} \}$. 
$LF(v)$ is the set of finite prefixes (or left 
factors) 
of the word $v$, and $LF(V)=\cup_{v\in V}LF(v)$ for every language $V$ of 
finite or 
infinite words. 

\hs We introduce now $\om$-context free languages via B\"uchi pushdown automata. 

\begin{Deff}\label{run}
A B\"uchi pushdown automaton is a 7-tuple $\mathcal{A}=(K,\Si,\Ga, \delta, q_0, Z_0, F)$,
where $K$ 
is a finite set of states, $\Sigma$ is a finite input alphabet, $\Gamma$ is
a 
finite pushdown alphabet,
$q_0\in K$ is the initial state, $Z_0 \in\Ga$ is the start symbol, $F \subseteq K$ is 
the set of final states,  
and $\delta$ is a mapping from $K \times (\Si\cup\{\lambda\} )\times \Ga $
to finite subsets of
$K\times \Gas$ . 
\nl
If $\gamma\in\Ga^{+}$ describes the pushdown store content, 
the leftmost symbol will be assumed to be on ``top" of the store.
A configuration of the BPDA $\mathcal{A}$ is a pair $(q, \gamma)$ where $q\in K$ and 
$\gamma\in\Gas$.
\nl
For $a\in \Si\cup\{\lambda\}$, $\gamma,\beta\in\Ga^{\star}$
and $Z\in\Ga$, if $(p,\beta)$ is in $\delta(q,a,Z)$, then we write
$a: (q,Z\gamma)\mapsto_\mathcal{A} (p,\beta\gamma)$.
\nl
Let $\sigma =a_1a_2\ldots a_n\ldots$ be an $\om$-word over $\Si$. 
A run of $\mathcal{A}$ on $\sigma$ is an 
infinite sequence $r=(q_i,\gamma_i, \varepsilon_i)_{i\geq 1}$ where 
$(q_i,\gamma_i)_{i\geq 1}$ is an infinite sequence of configurations of $\mathcal{A}$
and,  for all $i \geq 1$,  $\varepsilon_i \in \{0, 1\}$ and: 

\begin{enumerate}
\ite $(q_1,\gamma_1)=(q_0, Z_0)$
\ite for each $i\geq 1$, there exists $b_i\in\Si\cup\{\lambda\}$ 
satisfying 
\nl $b_i: (q_i,\gamma_i)\mapsto_\mathcal{A}(q_{i+1},\gamma_{i+1} )$
\nl and ( $\varepsilon_i=0$ iff $b_i=\lambda$ ) 
\nl and  such that  ~ $a_1a_2\ldots a_n\ldots =b_1b_2\ldots b_n\ldots$ 
\end{enumerate}

\noi $In(r)$ is the set of all states
entered infinitely often during run $r$.
\nl The \ol~ accepted by $\mathcal{A}$ is 
 $$L(\mathcal{A})= \{ \sigma\in\Si^\om \mid \mbox{  there exists a  run } r
\mbox{  of } \mathcal{A} \mbox{  on } 
\sigma \mbox{  such that } In(r) \cap F \neq\emptyset \}$$

\end{Deff}

\noi The class $CFL_\om$ of $\om$-context free languages is the class 
of \ol s accepted by B\"uchi  pushdown automata.  
It is also the $\om$-Kleene closure of the class $CFL$ of context free finitary 
languages, where  
for any family $\mathcal{L}$  of finitary \la s, the $\om$-Kleene closure 
of $\mathcal{L}$, is: ~~
$\om-KC(\mathcal{L}) = \{ \cup_{i=1}^n U_i.V_i^\om \mid \fa i\in [1, n]~~~ 
U_i, V_i \in \mathcal{L} \}$. 

\hs If we omit the pushdown stack and the $\lambda$-transitions, 
we get the classical notion of B\"uchi automaton. 
Recall that the class $REG_\om$ of \orl s is the 
class of \ol s accepted by finite automata with a B\"uchi acceptance 
condition. It is also the $\om$-Kleene closure of the class $REG$ of regular finitary 
languages, \cite{tho} \cite{sta} \cite{pp}.

\hs Notice that we introduced in the above definition the numbers $\varepsilon_i \in \{0, 1\}$ 
in order to distinguish runs of a BPDA which go through 
the same infinite sequence of configurations but 
for which $\lambda$-transitions do not occur 
at the same steps of the computations.  
We can now briefly recall  some definitions of \cite{ambcf} about ambiguity. 

\hs  We shall denote $\aleph_0$ the cardinal of $\om$, 
and $2^{\aleph_0}$ the cardinal of the continuum. It is 
also the cardinal of the set of real numbers and of the set 
$\Sio$ for every finite alphabet $\Si$ having at least two letters. 

\begin{Deff} 
Let $\mathcal{A}$ be a BPDA accepting infinite words over the alphabet $\Si$. 
For $x\in \Sio$ 
let $\alpha_\mathcal{A}(x)$ be the cardinal of the set of accepting runs of $\mathcal{A}$ on 
$x$. 
\end{Deff}

\begin{Lem}[\cite{ambcf}]\label{nb}
Let $\mathcal{A}$ be a BPDA accepting infinite words over the alphabet $\Si$. 
Then for all $x\in \Sio$ it holds that 
~~~~$\alpha_\mathcal{A}(x) \in \mathbb{N} \cup \{\aleph_0, 2^{\aleph_0}\}$.  
\end{Lem}

\begin{Deff} 
Let $\mathcal{A}$ be a BPDA accepting infinite words over the alphabet $\Si$. 
\begin{enumerate} 
\ite [(a)] If $\sup \{\alpha_\mathcal{A}(x) \mid  x\in \Sio \} \in \mathbb{N} \cup 
\{2^{\aleph_0}\}$, then 
$\alpha_\mathcal{A} = \sup \{\alpha_\mathcal{A}(x)   \mid   x\in \Sio \}$. 
\ite [(b)] If $\sup \{\alpha_\mathcal{A}(x)   \mid   x\in \Sio \} = \aleph_0$ and there 
is 
no word $x\in \Sio$ such that $\alpha_\mathcal{A}(x)=\aleph_0$, then 
$\alpha_\mathcal{A} = \aleph_0^-$. 
\nl ($\aleph_0^-$ does not represent a cardinal but is a new symbol that we 
introduce to conveniently speak of this situation). 
\ite [(c)] If $\sup \{\alpha_\mathcal{A}(x)   \mid   x\in \Sio \} = \aleph_0$ and there 
exists (at least) 
one word $x\in \Sio$ such that $\alpha_\mathcal{A}(x)=\aleph_0$, then 
$\alpha_\mathcal{A} = \aleph_0$ 
\end{enumerate} 
\end{Deff}

\noi Notice that for a BPDA $\mathcal{A}$, $\alpha_\mathcal{A}=0$ iff 
$\mathcal{A}$ does not accept any 
$\om$-word. 
\nl $\mathbb{N} \cup \{\aleph_0^-, \aleph_0, 2^{\aleph_0}\}$ 
is linearly ordered by the relation $<$ defined by 
$\forall k \in \mathbb{N}$, $k < k+1 < \aleph_0^- < \aleph_0 < 2^{\aleph_0}$.
\noi Now we can define a hierarchy  of $\om$-CFL: 

\begin{Deff} For  $k \in \mathbb{N} \cup \{\aleph_0^-, \aleph_0, 
2^{\aleph_0}\}$  let 
\nl $CFL_\om(\alpha \leq k) = \{L(\mathcal{A}) \mid  \mathcal{A} 
\mbox{ is a } BPDA \mbox{ with } 
\alpha_\mathcal{A} \leq k \}$  
\nl $CFL_\om(\alpha < k) = \{L(\mathcal{A}) \mid  \mathcal{A} 
\mbox{ is a } BPDA \mbox{ with } 
\alpha_\mathcal{A} < k \}$  
\nl $NA-CFL_\om = CFL_\om(\alpha \leq 1)$ is the class of non ambiguous 
$\om$-context free languages. 
\nl For every integer $k$ such that $k \geq 2$, or $k\in \{\aleph_0^-, 
\aleph_0, 2^{\aleph_0}\}$,
\nl $A(k)-CFL_\om = CFL_\om(\alpha \leq k) - CFL_\om(\alpha < k)$
\nl If $L \in A(k)-CFL_\om$ with $k \in \mathbb{N}, k \geq 2$, or $k\in \{\aleph_0^-, 
\aleph_0, 2^{\aleph_0}\}$, then $L$ is said to be inherently ambiguous 
of degree $k$. 
\end{Deff}

\noi Recall that one can define in a similar way the degree of ambiguity of 
a finitary 
context free language. If $M$ is a pushdown automaton 
 accepting finite words by final states (or by final 
states and 
topmost stack letter) then $\alpha_M 
\in \mathbb{N}$ 
or $\alpha_M=\aleph_0^-$ or $\alpha_M=\aleph_0$. 
However every context free language is accepted by a  pushdown automaton 
 $M$ with $\alpha_M \leq \aleph_0^-$, \cite{abb}. 
We shall denote, with similar notations as above, 
the class 
of non ambiguous context free languages by $NA-CFL$ and the class of 
inherently ambiguous 
context free languages  of degree $k\geq 2$ by $A(k)-CFL$. 
Then $A(\aleph_0^-)-CFL$ is usually 
called the class 
of context free languages  which are inherently ambiguous of infinite degree, \cite{her}. 

\hs Now we can state some  links between cases of finite and infinite 
words.

\begin{Pro}[\cite{ambcf}]\label{fin-inf-cf} 
Let $V\subseteq \Sis$ be a finitary context free language and $d$ be a new letter not in $\Si$, then 
the following equivalence holds for all 
$k \in \mathbb{N} \cup \{\aleph_0^-\}$: 
$$V.d^\om \mbox{ is in } CFL_\om(\alpha \leq k) \mbox{  iff } V \mbox{  is in  } 
CFL(\alpha \leq k)$$
\end{Pro}

\section{Borel and analytic sets}
\noi We assume the reader to be familiar with basic notions of topology 
which may be found in \cite{mos} \cite{lt} \cite{Ke} \cite{sta} \cite{pp}. 
\nl  For a  finite alphabet $X$ 
 we shall consider $X^\om$ as a topological space with the Cantor topology.
 The open sets of $X^\om$ are the sets in the form $W.X^\om$, where $W\subseteq X^\star$.
A set $L\subseteq X^\om$ is a closed set iff its complement $X^\om - L$ is an open set.
\nl  Define now the hierarchy of Borel sets of finite ranks:

\begin{Deff}
The classes ${\bf \Si_n^0}$ and ${\bf \Pi_n^0 }$ of the Borel hierarchy
 on the topological space $X^\om$  are defined as follows:
\nl ${\bf \Si^0_1 }$ is the class of open sets of $X^\om$.
\nl ${\bf \Pi^0_1 }$ is the class of closed sets of $X^\om$.
\nl And for any integer $n\geq 1$:
\nl ${\bf \Si^0_{n+1} }$   is the class of countable unions 
of ${\bf \Pi^0_n }$-subsets of  $X^\om$.
\nl ${\bf \Pi^0_{n+1} }$ is the class of countable intersections of 
${\bf \Si^0_n}$-subsets of $X^\om$.

\end{Deff}

\noi  The Borel hierarchy is also defined for transfinite levels, but we shall not 
need them in the present study. The class of Borel subsets of $X^\om$  is the closure 
of the class of open subsets of $X^\om$  under complementation and countable unions (hence 
also under countable intersections) 
There are also some subsets of $X^\om$ which are not Borel.  
In particular the class of Borel subsets of $X^\om$ is strictly included into 
the class  ${\bf \Si^1_1}$ of analytic sets which are 
obtained by projection of Borel sets.  
\nl Notice that if $\Si$ and $\Ga$ are two finite alphabets then the product $\Sio \times \Gao$ 
can be identified with the space $(\Si \times \Ga)^\om$  and we always consider in the sequel 
that such a space $\Sio \times \Gao$ is equipped with the Cantor topology. 

\begin{Deff} A set $A\subseteq \Sio$ is an analytic set if there is a finite alphabet 
$\Ga$ and a Borel set 
$B\subseteq \Sio \times \Gao$ such that 
 $A = \{\alpha \in \Sio \mid 
\exists \beta \in \Gao~~(\alpha,\beta) \in B\}$. 
\nl A set $C\subseteq \Sio $ is coanalytic if its complement $\Sio - C$ is analytic.
The class of analytic sets is denoted ${\bf \Si^1_1}$ and the class of 
coanalytic sets is denoted ${\bf \Pi^1_1}$. 
\end{Deff}

\hs Recall also the notion of completeness with regard to reduction by continuous functions. 
For an integer $n\geq 1$, a set $F\subseteq X^\om$ is said to be 
a ${\bf \Si^0_n}$  (respectively,  ${\bf \Pi^0_n}$, ${\bf \Si^1_1}$, 
${\bf \Pi^1_1}$)-complete set 
iff for any set $E\subseteq Y^\om$  (with $Y$ a finite alphabet): 
 $E\in {\bf \Si^0_n}$ (respectively,  $E\in {\bf \Pi^0_n}$,  
$E\in {\bf \Si^1_1}$, $E\in {\bf \Pi^1_1}$) 
iff there exists a continuous function $f: Y^\om \ra X^\om$ such that $E = f^{-1}(F)$.  
\nl ${\bf \Si^0_n}$
 (respectively,  ${\bf \Pi^0_n}$)-complete sets, with $n$ an integer $\geq 1$, 
 are thoroughly characterized in \cite{stac}.

\section{Topology and ambiguity in  $\om$-context free languages}

\noi Let $\Si$ and $X$ be two finite alphabets. 
If $B \subseteq \Sio  \times  X^\om $ and $\alpha 
\in \Sio$, the section in $\alpha$ of $B$ is 
 $B_\alpha=\{\beta \in X^\om \mid (\alpha,\beta) \in B\}$ and   
the  projection of $B$ on $\Sio$ is the set 
$PROJ_{\Sio}(B)=\{\alpha \in \Sio \mid B_\alpha \neq \emptyset \} = 
\{\alpha \in \Sio \mid  \exists \beta ~(\alpha, \beta) \in B\}$. 

\hs  We are going to prove the following lemma which will be useful in the sequel: 

\begin{Lem}\label{cor2}
Let $\Si$ and $X$ be two finite alphabets having at least two letters and 
 $B$ be a Borel subset of 
$\Sio \times X^\om$ such that $PROJ_{\Sio}(B)$ is not a Borel subset of $\Sio$.
Then there are $2^{\aleph_0}$ $\om$-words  $\alpha \in \Sio$ such that the section $B_\alpha$ 
has cardinality $2^{\aleph_0}$.
\end{Lem}

\proo Let $\Si$ and $X$ be two finite alphabets having at least two letters and 
 $B$ be a Borel subset of 
$\Sio \times X^\om$ such that $PROJ_{\Sio}(B)$ is not  Borel.  

\hs In a first step we shall prove that 
there are uncountably many $\alpha \in \Sio$ such that the section $B_\alpha$ 
is uncountable.

\hs Recall that by a Theorem of Lusin and Novikov, see \cite[page 123]{Ke}, if for
 all  $\alpha \in \Sio$, the section $B_\alpha$ of the Borel set $B$ was   countable,  
 then  $PROJ_{\Sio}(B)$ would be  a Borel subset of $\Sio$. 

\hs Thus there exists at least  one $\alpha \in \Sio$ such that  $B_\alpha$  is uncountable.
In fact we have not only one  $\alpha$ such that $B_\alpha$  is uncountable.  

\hs For $\alpha \in \Sio$ we have 
$\{\alpha\} \times B_\alpha = B \cap [ \{\alpha\} \times X^\om ]$. 
But $\{\alpha\} \times X^\om$ is a closed 
hence Borel subset of $\Sio \times X^\om$ thus $\{\alpha\} \times B_\alpha$ 
is Borel as intersection of two Borel sets.

\hs If there was only one $\alpha \in \Sio$ such that  $B_\alpha$  is uncountable, then 
$C=\{\alpha\}\times B_\alpha$ would be Borel so $D=B - C$ would be borel 
because the class of Borel sets is closed 
under boolean operations. 
\nl But all sections of $D$ would be countable thus 
 $PROJ_{\Sio}(D)$ would be Borel by Lusin and Novikov's Theorem. 
Then $PROJ_{\Sio}(B)= \{\alpha\}\cup PROJ_{\Sio}(D)$ 
would be also Borel as union of two Borel sets, and this would lead to a
contradiction. 

\hs In a similar manner we can prove that  the set $U=\{\alpha \in \Sio \mid  B_\alpha  
\mbox{ is uncountable } \}$ is uncountable, otherwise 
$U=\{\alpha_0, \alpha_1, \ldots \alpha_n, \ldots \}$ would be Borel as the countable union 
of the closed sets $\{\alpha_i\}$, $i\geq 0$. 
\nl For each $n\geq 0$ the set $\{\alpha_n\}\times B_{\alpha_n}$ would be Borel,  
and 
$C=\cup_{n \in \om}\{\alpha_n\}\times B_{\alpha_n}$ would be Borel as a 
countable union of Borel sets. 
So $D=B - C$ would be borel too. 
\nl But all sections of $D$ would be countable thus 
 $PROJ_{\Sio}(D)$ would be Borel by Lusin and Novikov's Theorem. 
 Then $PROJ_{\Sio}(B)= U \cup PROJ_{\Sio}(D)$ would be 
also Borel as union of two Borel sets, and this would lead to a
contradiction. 

\hs So we have proved that the set 
$\{ \alpha \in \Sio \mid B_\alpha \mbox{ is uncountable } \}$
 is uncountable. 

\hs On the other hand we know from 
another Theorem of Descriptive Set Theory that the set 
$\{ \alpha \in \Sio \mid B_\alpha \mbox{ is countable } \}$ is a ${\bf \Pi^1_1}$-subset of 
$\Sio$, see \cite[page 123]{Ke}.  
\nl Thus its complement $\{ \alpha \in \Sio \mid B_\alpha \mbox{ is uncountable } \}$
is analytic. 
But by Suslin's Theorem an analytic subset of   $\Sio$ is either countable 
or has cardinality $2^{\aleph_0}$, \cite[p. 88]{Ke}. Therefore the set 
$\{ \alpha \in \Sio \mid B_\alpha \mbox{ is uncountable } \}$
has cardinality $2^{\aleph_0}$.

\hs Recall now  that we have already seen that, for each $\alpha \in \Sio$, the set 
 $\{\alpha\} \times B_\alpha$ is Borel. 
We can then infer that $B_\alpha$ 
itself is Borel by considering the function $h: X^\om \ra \Sio \times X^\om$ defined by 
$h(\sigma)=(\alpha, \sigma)$ for all $\sigma \in X^\om$.  The function $h$ is continuous 
and $B_\alpha = h^{-1}(\{\alpha\} \times B_\alpha)$. So  $B_\alpha$ is Borel because 
the inverse image of a Borel set by a continuous function is a Borel set. 
Again by Suslin's Theorem $B_\alpha$ is either countable or has cardinality $2^{\aleph_0}$.
From this we deduce that 
$\{ \alpha \in \Sio \mid B_\alpha \mbox{ is uncountable } \} = 
\{ \alpha \in \Sio \mid B_\alpha \mbox{ has cardinality } 2^{\aleph_0} \}$ 
has cardinality $2^{\aleph_0}$. \ep

\hs  We can now infer some results for $\om$-context free languages. 

\begin{The}\label{mainthe}
Let $L(\mathcal{A})$ be an $\om$-CFL accepted by a BPDA $\mathcal{A}$ such that $L(\mathcal{A})$ 
is an analytic but non Borel set. The set of $\om$-words, 
which have $2^{\aleph_0}$ accepting runs by $\mathcal{A}$, has cardinality $2^{\aleph_0}$. 
\end{The}

\proo   Let  
$\mathcal{A}=(K,\Si,\Ga, \delta, q_0, Z_0, F)$ be a BPDA such that $L(\mathcal{A})$ 
is an analytic but non Borel set. 

\hs To an infinite sequence $r=(q_i,\gamma_i, \varepsilon_i)_{i\geq 1}$, where 
for all  $i\geq 1$,  $q_i \in K$, $\gamma_i \in \Ga^+$ and 
$\varepsilon_i \in\{0,1\}$, 
we associate 
 an $\om$-word $\bar{r}$ over the alphabet $X = \Ga\cup K \cup\{0, 1\}$ defined by 
$$\bar{r}=q_1.\gamma_1.\varepsilon_1.q_2.\gamma_2.\varepsilon_2 
\ldots q_i.\gamma_i.\varepsilon_i \ldots$$
Then to an infinite word $\sigma\in \Sio$ and 
an infinite sequence $r=(q_i,\gamma_i, \varepsilon_i)_{i\geq1}$, we
associate 
the couple  $(\sigma, \bar{r}) \in \Sio
\times (\Ga \cup K \cup\{0, 1\})^\om$. 

\hs Recall now that ${\bf \Pi^0_2 }$-subsets of a Cantor set $\Si^\om$  
are characterized in the following way. 
For $W\subseteq \Si^\star$ the $\delta$-limit $W^\delta$ of $W$ is 
the set of $\om$-words over $\Si$ having 
infinitely many prefixes in $W$: ~  
$W^\delta=\{\sigma\in \Si^\om \mid  \exists^\om i \mbox{ such that }
\sigma(1)\ldots \sigma(i)\in W\}$. 
 Then a subset $L$ of $\Si^\om$ is a ${\bf \Pi^0_2 }$-subset of $\Si^\om$ iff there
exists
a set $W\subseteq \Si^\star$ such that $L=W^\delta$,  \cite{sta} \cite{pp}.  

\hs It is then easy to see that the set 
$$R = \{ (\sigma,  \bar{r}) \mid \bar{r} \mbox{ is the code of an accepting run of } 
\mathcal{A}    \mbox{ over } \sigma \}$$ 
\noi is a ${\bf \Pi^0_2}$-subset of 
$\Sio \times X^\om = (\Si \times X)^\om$ as intersection of two ${\bf \Pi^0_2}$-sets. 
In fact we have 
$R=(R')^\delta \cap (R'')^\delta $ where 
$R'\subseteq (\Si \times X)^+$ is the set of couples of words 
$(u, v)$ in the form: 
$$u = a_1.a_2. \ldots a_p$$
$$v = q_1.\gamma_1.\varepsilon_1.q_2.\gamma_2.\varepsilon_2 
\ldots q_n.\gamma_n.\varepsilon_n$$
\noi where for each $i\in [1, p]~~ a_i \in \Si$, for each 
$i\in [1, n]~~ q_i \in K$, $\gamma_i \in \Ga^+$ and $\varepsilon_i \in \{0, 1\}$. 
Moreover $|u|=|v|$, $\varepsilon_n=1$, and

\begin{enumerate}
\ite $(q_1,\gamma_1)=(q_0, Z_0)$
\ite for each $i \in [1, n-1]$, there exists $b_i\in\Si\cup\{\lambda\}$ 
satisfying 
\nl $b_i: (q_i,\gamma_i)\mapsto_\mathcal{A}(q_{i+1},\gamma_{i+1} )$
\nl and ( $\varepsilon_i=0$ iff $b_i=\lambda$ ) 
\nl and  such that  ~ $b_1b_2\ldots b_{n-1}$ is a prefix of $u = a_1.a_2. \ldots a_p$. 
\end{enumerate}

\noi And $R''\subseteq (\Si \times X)^+$ is the set of couples of words 
$(u, v) \in \Si^+ \times X^+$ such that $|u|=|v|$ and 
the last letter of $v$ is an element $q \in F$. 

\hs  
In particular $R$  is a Borel subset of $\Sio \times X^\om$. 
But by definition of $R$ it turns out that $PROJ_{\Sio}(R)=L(\mathcal{A})$ so 
$PROJ_{\Sio}(R)$ is not Borel. Thus Lemma \ref{cor2} implies that  
there are $2^{\aleph_0}$ $\om$-words  $\alpha \in \Sio$ such that $R_\alpha$ has cardinality 
$2^{\aleph_0}$. This means that these words have $2^{\aleph_0}$ 
accepting runs by the B\"uchi pushdown automaton $\mathcal{A}$. 
\ep

\begin{Exa}\label{max-amb}
Let $\Si=\{0,1\}$ and  $d$ be a new letter not in $\Si$ and 
$$D=\{ u.d.v  \mid  u, v \in \Sis ~and~ ( |v|=2|u|)~~ or ~~( |v|=2|u|+1)~ 
\}$$ 
\noi $D\subseteq (\Si \cup \{d\})^\star$ is a context free language. 
Let $g:\Si\ra \mathcal{P}((\Si \cup \{d\})^\star)$ be the substitution 
defined by 
$g(a)=a.D$. As $W=0^\star1$ is regular, 
$g(W)$ is a context free language, thus 
$(g(W))^\om$ is an $\om$-CFL. It is proved in \cite{finb} that $(g(W))^\om$ is 
${\bf \Si^1_1}$-complete. In particular $(g(W))^\om$ is an analytic non Borel set. 
Thus every BPDA accepting 
$(g(W))^\om$ has the maximum ambiguity and $(g(W))^\om \in A(2^{\aleph_0})-CFL_\om$. 

\hs On the other hand 
we can  prove that $g(W)$ is a non ambiguous context free language. 

\hs For that purpose consider a (finite) word $x\in g(W)$;
 then $x\in g(0^n.1)$ for some integer $n\geq 0$. Therefore $x$ may 
be written in the form

$$x = 0.u_1.d.v_1.0.u_2.d.v_2 \ldots 0.u_n.d.v_n.1.u_{n+1}.d.v_{n+1}$$

\noi where $u_i.d.v_i \in D$ holds for all $i\in [1, n+1]$. 
It is easy to see that 
the length $|v_{n+1}|$ and the word $v_{n+1}$ are determined by the word $x$:  
$v_{n+1}$ is the suffix of $x$ following the last letter $d$ of $x$, and  
 $|v_{n+1}|=2|u_{n+1}|$ (if $|v_{n+1}|$ is even) or  $|v_{n+1}|=2|u_{n+1}| + 1$ 
(if $|v_{n+1}|$ is odd) thus  
$|u_{n+1}|$ is determined by $|v_{n+1}|$ hence $u_{n+1}$ is also determined. 
 Next one can see that  $v_n$ also is
fixed by $x$ (the word $v_n.1.u_{n+1}$ is the segment of $x$ which is located between the 
$n^{th}$ and the $(n+1)^{th}$ occurrences of the letter $d$ in $x$ and knowing $u_{n+1}$ 
gives us $v_n$). 
\nl We can similarly prove by induction on the integer $k$ that the words 
$v_{n+1-k}$ and $u_{n+1-k}$, for $k \in [0, n]$, are uniquely determined by $x$. 
\nl Therefore the  word $x$ admits a unique decomposition in the above form. We can 
then easily construct a pushdown automaton (and even a one counter automaton) which 
accepts the language $g(W)$ and which is non ambiguous. So the language 
$g(W)$ is a non ambiguous context free language. 
\end{Exa}

\noi The above example shows that the $\om$-power of a non ambiguous context free language 
may have maximum ambiguity. Conversely consider the context free 
language $V = V_1\cup V_2 \subseteq \{a, b, c\}^\star$ where 
$V_1 = \{a^nb^nc^p \mid  n\geq 1,~ p\geq 1 \}$ and 
$V_2 = \{a^nb^pc^p \mid  n\geq 1,~ p\geq 1 \}$. 
$V_1$ and $V_2$ are deterministic context free, hence they are 
non ambiguous  context free languages. 
But their union $V$ is an inherently ambiguous context free 
language \cite{mau}. $V^\star$ is a context free language 
which is inherently ambiguous of infinite degree 
(and it is proved in \cite{naj} that it is 
even exponentially ambiguous in the sense 
of Naji and Wich, see also  \cite{wich1} about this notion). 
Let then $L= V^\star \cup \{a, b, c\}$. 
The language $L$ is still a context free language 
which is inherently ambiguous of infinite degree and $L^\om = \{a, b, c \}^\om$ 
is an \orl~ hence it is a non ambiguous $\omega$-context free language.

\hs We have then proved that neither unambiguity nor inherent ambiguity is preserved 
by the operation $L \ra L^\om$:

\hs 

\begin{Pro}
\noi 
\begin{enumerate}
\ite  There exists a non ambiguous context free finitary language $L$ such that 
$L^\om$ is in $A(2^{\aleph_0})-CFL_\om$. 
\ite  There exists a context free finitary language $L$, which is inherently ambiguous of 
infinite degree, such that $L^\om$ is a non ambiguous $\om$-context free language. 
\end{enumerate}
\end{Pro}

\noi We can also consider the above mentioned language $g(W)$ in the context of code theory. 
We have proved  that $g(W)$ is a non ambiguous context free language.  By a 
similar reasoning we can prove that $g(W)$ is a code, i.e. that every (finite) word 
$y\in g(W)^+$ has a unique decomposition $y=x_1.x_2 \ldots x_n$ in words $x_i \in g(W)$. 
\nl On the other side $g(W)$ is not an $\om$-code, i.e. some words  $z \in g(W)^\om$ have 
several  decompositions in the form $z = x_1.x_2 \ldots x_n \ldots $ where 
for all $i\geq 1$~ $x_i \in g(W)$. In fact we can get a much stronger result, using 
Lemma \ref{cor2}: 

\begin{Fact}
There are $2^{\aleph_0}$ $\om$-words in  $g(W)^\om$  which have 
 $2^{\aleph_0}$  decompositions  in words in $g(W)$. 
\end{Fact}

\proo  We can fix a recursive  enumeration $\theta$ of the set $g(W)$. 
So the function $\theta : \mathbb{N} \ra g(W)$ is a bijection  and we denote $u_i=\theta(i)$. 
\nl Let now $\mathcal{D}$ be the set of couples 
$(\sigma, x) \in \{0, 1\}^\om \times (\Si \cup \{d\})^\om$ such that:  

\begin{enumerate} 
\ite $\sigma \in (0^\star.1)^\om$, so $\sigma$ may be written in the form 
$$\sigma = 0^{n_1}.1.0^{n_2}.1.0^{n_3}.1 \ldots 0^{n_p}.1.0^{n_{p+1}}.1 \ldots $$
\noi where $\fa i \geq 1$ ~ $n_i \geq 0$, ~and  
\ite $$x = u_{n_1}.u_{n_2}.u_{n_3}\ldots u_{n_p}.u_{n_{p+1}} \ldots $$
\end{enumerate} 

\noi $\mathcal{D}$ is a Borel subset of  $\{0, 1\}^\om \times (\Si \cup \{d\})^\om$ 
because it is accepted by a deterministic Turing machine with a B\"uchi acceptance condition 
\cite{sta}. 
On the other hand  $PROJ_{(\Si \cup \{d\})^\om}(\mathcal{D})=g(W)^\om$ is not Borel and  
Lemma \ref{cor2}  implies that there are $2^{\aleph_0}$ $\om$-words $x$ in  $ g(W)^\om$  
such that $\mathcal{D}_x$ has cardinality $2^{\aleph_0}$. This means that
 there are $2^{\aleph_0}$ $\om$-words  $x\in g(W)^\om$ which have 
 $2^{\aleph_0}$  decompositions  in words in $g(W)$. 
\nl We can say that the code $g(W)$ is really not an $\om$-code !  \ep

\hs The result given by Theorem \ref{mainthe}  may be compared with 
a general study of topological properties of transition systems 
due to Arnold 
\cite{arn2}. If we consider a 
BPDA as a transition system with infinitely many states, Arnold's results 
imply 
that every non ambiguous $\om$-CFL is a Borel set. On the other side 
\de $\om$-CFL have not a great topological complexity, because they are 
boolean combinations 
of ${\bf \Pi_2^0}$-sets. We know some examples of non ambiguous $\om$-CFL of 
every finite 
Borel rank, but none of infinite Borel rank. 
These results led the first author  to the following question: are there some more links 
between the topological complexity of an $\om$-CFL and the ambiguity of BPDA which 
accept it? 
In \cite{ambcf}  the well known notions 
of degrees of ambiguity for CFL are extended to $\om$-CFL 
and such  supposed 
connections are investigated.  In particular, 
using results of Duparc on the Wadge hierarchy, which is a 
great refinement of the Borel hierarchy \cite{dup}, it is proved 
 that 
for each $k$ such that $k$ is an integer $\geq 2$ or 
$k=\aleph_0^-$ and for  each integer $n\geq 1$, there exist in $A(k)-CFL_\om$ some 
${\bf \Si_n^0}$-complete $\om$-CFL 
and some ${\bf \Pi_n^0}$-complete $\om$-CFL.

\noi In the proofs of these results is  used 
the operation $W\ra Adh(W)$ where for a finitary language $W\subseteq \Si^\star$,   
$Adh(W)=\{ \sigma \in \Si^\om \mid  LF(\sigma) \subseteq LF(W) \}$
is the  adherence  of $W$. 
 We recall that 
a set $L\subseteq \Si^\om$ is a closed set of $\Si^\om$ iff there exists a
finitary
language $W\subseteq \Si^\star$ such that $L=Adh(W)$.
\nl  It is well known that if $W$ is a context free language, then $Adh(W)$
is in $CFL_\om$. Moreover every closed (\de) $\om$-CFL is the adherence of a
(\de) context free language, \cite{sta}. 
\nl So the question of the preservation of ambiguity by the operation $W\ra Adh(W)$ 
naturally arises. 

\begin{Pro}
Neither unambiguity nor inherent ambiguity is preserved by taking the
adherence   of a finitary context free language.
\end{Pro} 

\proo (I) We are firstly  looking for a non ambiguous finitary context free
language  which have an inherently ambiguous adherence. Let then the
following finitary language over the alphabet
$\{a, b, c, d\}$:
$$L_1 = \{a^nb^nc^p.d^{2i} \mid  n, p, i \mbox{ are integers } \geq 1 \}
\cup \{a^nb^pc^p.d^{2i+1}  \mid  n, p, i \mbox{ are integers } \geq 1 \}$$

\noi $L_1$ is the disjoint union of two \de (hence non ambiguous)
finitary context free languages thus it is a non ambiguous CFL because the
class
$NA-CFL$ is closed under finite disjoint union. 
It is easy to see that the adherence of $L_1$ is

$$Adh(L_1) = \{a^\om\} \bigcup a^+.b^\om \bigcup
\{a^nb^n  \mid   n \geq 1 \}.c^\om \bigcup (V_1 \cup V_2).d^\om$$

\noi where $V_1 = \{a^nb^nc^p \mid  n\geq 1,~ p\geq 1 \}$ and 
$V_2 = \{a^nb^pc^p \mid  n\geq 1,~ p\geq 1 \}$. 
Then it holds that
$Adh(L_1) \cap a^+.b^+.c^+.d^\om = (V_1 \cup V_2).d^\om = V.d^\om$, where $V=V_1 \cup V_2$. 
\nl By proposition \ref{fin-inf-cf}, the $\omega$-context free language  $V.d^\om$ is
inherently ambiguous 
because  $V$ is inherently ambiguous \cite{mau}.
Thus $Adh(L_1)$ is inherently ambiguous because otherwise
$V.d^\om$ would be non ambiguous because the class $NA-CFL_\om$ is closed
under
intersection with
\orl s \cite{ambcf}, and  $a^+.b^+.c^+.d^\om$ is an \orl .  

\hs (II) We are now looking for an inherently ambiguous context free language  
which have a non ambiguous adherence. We shall use a result of Crestin,  \cite{cre}:  
the language $C=\{u.v  \mid   u, v \in \{a, b\}^+ \mbox{ and } u^R=u
\mbox{ and } v^R=v \}$ is a context free language which is inherently
ambiguous (of infinite degree). 
 In fact $C=L_p^2$ where $L_p=\{ v \in \{a, b\}^+  \mid   v^R=v \}$ is the
language of palindromes over the alphabet $\{a, b\}$.
Consider now the adherence of the language $C$.
$Adh(C) = \{a, b\}^\om$
holds because every word $u\in \{a, b\}^\star $ is a prefix of a palindrome
(for example of the palindrome $u.u^R$) hence it is also a prefix of a
word of $C$.
Thus $C$ is inherently ambiguous and $Adh(C)$ is a non ambiguous
$\om$-context free language because it is an \orl .
\ep

\hs We have seen that  closed sets are characterized as
 adherences of  finitary languages. Similarly  we have already seen, 
in the proof of Theorem \ref{mainthe},  that 
${\bf \Pi^0_2 }$-subsets of $\Si^\om$  
are characterized as $\delta$-limits $W^\delta$ of finitary languages $W\subseteq \Sis$. 
\nl Recall that $W \in REG$ implies that $W^\delta \in REG_\om$. But there exist
some
context free languages $L$ such that $L^\delta$ is not in $CFL_\om$; see
\cite{sta}
for an example of such a language $L$. In the case $W\in CFL$ {\bf and}
$W^\delta \in CFL_\om$,
the question naturally arises of the preservation of ambiguity by the
operation
$W\ra W^\delta$. The answer is given by the following:

\begin{Pro}
Neither unambiguity nor inherent ambiguity is preserved by taking the
$\delta$-limit
of a finitary context free language.
\end{Pro}

\proo (I) Let again $L_1$ be the
following finitary language over the alphabet
$\{a, b, c, d\}$:
$$L_1 = \{a^nb^nc^p.d^{2i}  \mid   n, p, i \mbox{ are integers } \geq 1 \}
\cup \{a^nb^pc^p.d^{2i+1}  \mid   n, p, i \mbox{ are integers } \geq 1 \}$$

\noi $L_1$ is
a non ambiguous CFL.
And the $\delta$-limit of the language $L_1$ is
$(L_1)^\delta = (V_1 \cup V_2).d^\om = V.d^\om$. 
 We have already seen that this \ol~ is an inherently ambiguous
$\om$-CFL. 

\hs (II) Consider now the inherently ambiguous context free language
$V  = \{a^nb^nc^p  \mid   n, p \geq 1 \}
\cup \{a^nb^pc^p  \mid   n, p \geq 1 \}$. 
Its $\delta$-limit is
$V^\delta = \{a^n.b^n  \mid   n \geq 1 \}.c^\om$. 
It is easy to see that $V^\delta$ is a \de $\om$-CFL hence it is a
non ambiguous $\om$-CFL. 
\ep

\section{Topology and ambiguity in  
infinitary rational relations}     

\noi 
Infinitary rational relations 
are subsets of  $\Sio \times \Gao$,  where 
$\Si$ and  $\Ga$ are finite alphabets, which are accepted by 
$2$-tape B\"uchi automata. 
\nl We are going to see in this section that some above methods can also be used 
in the case of infinitary rational relations. 

\begin{Deff}
A  2-tape B\"uchi automaton (2-BA)
 is a sextuple $\mathcal{T}=(K, \Si, \Ga, \Delta, q_0, F)$, where 
$K$ is a finite set of states, $\Si$ and $\Ga$ are finite  alphabets, 
$\Delta$ is a finite subset of $K \times \Sis \times \Gas \times K$ called 
the set of transitions, $q_0$ is the initial state,  and $F \subseteq K$ is the set of 
accepting states. 
\nl A computation $\mathcal{C}$ of the  
2-tape B\"uchi automaton $\mathcal{T}$ is an infinite sequence of transitions 
$$(q_0, u_1, v_1, q_1), (q_1, u_2, v_2, q_2), \ldots (q_{i-1}, u_{i}, v_{i}, q_{i}), 
(q_i, u_{i+1}, v_{i+1}, q_{i+1}), \ldots $$
\noi The computation is said to be successful iff there exists a final state $q_f \in F$ 
and infinitely many integers $i\geq 0$ such that $q_i=q_f$. 
\nl The input word of the computation is $u=u_1.u_2.u_3 \ldots$
\nl The output word of the computation is $v=v_1.v_2.v_3 \ldots$
\nl Then the input and the output words may be finite or infinite. 
\nl The infinitary rational relation $R(\mathcal{T})\subseteq \Sio \times \Ga^\om$ 
accepted by the 2-tape B\"uchi automaton $\mathcal{T}$ 
is the set of couples $(u, v) \in \Sio \times \Ga^\om$ such that $u$ and $v$ are the input 
and the output words of some successful computation $\mathcal{C}$ of $\mathcal{T}$. 
\nl The set of infinitary rational relations will be denoted $RAT_\om$. 
\end{Deff} 

\noi One can define degrees of ambiguity for 2-tape B\"uchi automata and for 
infinitary rational relations as in the case of BPDA and $\om$-CFL.

\begin{Deff} 
Let $\mathcal{T}$ be a 2-BA accepting couples of infinite words 
of $\Sio \times \Ga^\om$.  For $(u, v) \in \Sio \times \Ga^\om$, let 
 $\alpha_{\mathcal{T}}(u, v)$ be   the cardinal of 
the set of accepting computations of $\mathcal{T}$ on $(u, v)$. 
\end{Deff} 

\begin{Lem}
Let $\mathcal{T}$ be a 2-BA accepting couples of infinite words 
 $(u, v) \in \Sio \times \Ga^\om$.  
Then for all $(u, v) \in \Sio \times \Ga^\om$ it holds that 
~~~~$\alpha_{\mathcal{T}}(u, v)  \in \mathbb{N} \cup \{\aleph_0, 2^{\aleph_0}\}$.  
\end{Lem}

\noi 
The proof that a value between $\aleph_0$ and $2^{\aleph_0}$ is impossible   
 follows from Suslin's Theorem because one can obtain the set of  codes  of 
accepting computations of $\mathcal{T}$ on $(u, v)$ as a section of a Borel set  
(see proof of next theorem) 
hence as a Borel  set. A similar reasoning was used  in the proof 
of Lemma \ref{nb}, \cite{ambcf}.

\begin{Deff} 
Let $\mathcal{T}$ be a 2-BA accepting couples of infinite words 
 $(u, v) \in \Sio \times \Ga^\om$.  
\begin{enumerate} 
\ite [(a)] If $\sup \{   \alpha_{\mathcal{T}}(u, v)   \mid  
(u, v) \in \Sio \times \Ga^\om \} \in \mathbb{N} \cup 
\{2^{\aleph_0}\}$, then 
$ \alpha_{\mathcal{T}} = \sup \{   \alpha_{\mathcal{T}}(u, v)   \mid  
(u, v) \in \Sio \times \Ga^\om \}$. 
\ite [(b)] If $\sup \{   \alpha_{\mathcal{T}}(u, v)   \mid  
(u, v) \in \Sio \times \Ga^\om \} = \aleph_0$ and there 
is no $(u, v) \in \Sio \times \Ga^\om$  such that 
$\alpha_{\mathcal{T}}(u, v) =\aleph_0$, then 
$\alpha_{\mathcal{T}}  = \aleph_0^-$. 
\ite [(c)] If $\sup \{   \alpha_{\mathcal{T}}(u, v)   \mid  
(u, v) \in \Sio \times \Ga^\om \} = \aleph_0$ and there 
exists (at least) 
one couple $(u, v) \in \Sio \times \Ga^\om$  such that 
$\alpha_{\mathcal{T}}(u, v) =\aleph_0$, then 
$ \alpha_{\mathcal{T}}  = \aleph_0$ 
\end{enumerate} 
\end{Deff}

\noi The set  $\mathbb{N} \cup \{\aleph_0^-, \aleph_0, 2^{\aleph_0}\}$ 
is linearly ordered as above by the relation $<$.

\begin{Deff} For  $k \in \mathbb{N} \cup \{\aleph_0^-, \aleph_0, 
2^{\aleph_0}\}$, let 
\nl $RAT_\om(\alpha \leq k) = \{  R(\mathcal{T}) \mid \mathcal{T} 
\mbox{ is a } 2-BA \mbox{ with } 
\alpha_{\mathcal{T}}   \leq k \}$ 
\nl $RAT_\om(\alpha < k) = \{  R(\mathcal{T}) \mid \mathcal{T} 
\mbox{ is a } 2-BA \mbox{ with } 
\alpha_{\mathcal{T}}  < k \}$ 
\nl $NA-RAT_\om = RAT_\om(\alpha \leq 1)$ is the class of non ambiguous 
infinitary rational relations. 
\nl For every integer $k \geq 2$, or $k\in \{\aleph_0^-, 
\aleph_0, 2^{\aleph_0}\}$,
\nl $A(k)-RAT_\om = RAT_\om(\alpha \leq k) - RAT_\om(\alpha < k)$ is the class of 
infinitary rational relations which are inherently ambiguous 
of degree $k$. 
\end{Deff}

\noi As for $\om$-context free languages, one can use Lemma \ref{cor2} 
to prove the following result. 

\begin{The}\label{mainthe2}
Let $R(\mathcal{T})\subseteq \Sio \times \Ga^\om$ be an infinitary rational relation 
accepted by a 2-tape B\"uchi automaton $\mathcal{T}$ such that $R(\mathcal{T})$
is an analytic but non Borel set. The set of couples of $\om$-words, 
which have $2^{\aleph_0}$ accepting computations  by  $\mathcal{T}$, 
has cardinality $2^{\aleph_0}$. 
\end{The}

\proo  It is very similar to proof of Theorem \ref{mainthe}.  
Let $R(\mathcal{T})\subseteq \Sio \times \Ga^\om$ be an infinitary rational relation 
accepted by a 2-tape B\"uchi automaton $\mathcal{T}=(K, \Si, \Ga, \Delta, q_0, F)$.  
We assume also that $R(\mathcal{T})$ is an analytic but non Borel set. 
  To an infinite sequence 
$$\mathcal{C} = 
(q_0, u_1, v_1, q_1), (q_1, u_2, v_2, q_2), \ldots (q_{i-1}, u_{i}, v_{i}, q_{i}), 
(q_i, u_{i+1}, v_{i+1}, q_{i+1}), \ldots $$
\noi where 
for all  $i\geq 0$,  $q_i \in K$, for all  $i\geq 1$,  $u_i \in \Sis$ and $v_i \in \Gas$,  
we associate 
an $\om$-word $\bar{\mathcal{C}}$ over the alphabet $X =  K \cup \Si \cup \Ga \cup\{e\}$, where 
$e$ is an additional letter.   $\bar{\mathcal{C}}$ is 
defined by: 
$$ \bar{\mathcal{C}}  =q_0.u_1.e.v_1.q_1.u_2.e.v_2.q_2
\ldots q_i.u_{i+1}.e.v_{i+1}.q_{i+1} \ldots$$
\noi Then the set
$$ \{ (u, v, \bar{\mathcal{C}}) \in \Sio \times \Ga^\om \times X^\om \mid \bar{\mathcal{C}} 
\mbox{ is the code of an accepting computation  of } \mathcal{T} 
 \mbox{ over } (u, v) \}$$ 
\noi is accepted by a deterministic Turing machine with a B\"uchi acceptance condition 
thus it is a ${\bf \Pi^0_2}$-set. 
 We can conclude as in proof of Theorem \ref{mainthe}. \ep 

\hs The first author showed that there exist 
some ${\bf \Si^1_1}$-complete, hence non Borel,  infinitary rational relations \cite{relrat}. 
So we can deduce the following result.  

\begin{Cor}\label{cor-relrat}
There exist some infinitary rational relations which are inherently ambiguous of 
degree  $2^{\aleph_0}$. 
\end{Cor}

\begin{Rem} Looking carefully at the example of non Borel infinitary rational relation 
given in \cite{relrat}, we can find a rational relation $S$ over finite words such that 
$S$ is non ambiguous and $S^\om$ is non Borel. So $S$ is a finitary rational relation 
which is non ambiguous but  $S^\om$ has maximum ambiguity because 
$S^\om \in A(2^{\aleph_0})-RAT_\om$ 
holds by Theorem \ref{mainthe2}. 
\end{Rem}

\noi Moreover the question of the decidability of ambiguity for infinitary rational relations 
naturally arises. It can  be solved, using another recent result of the first author. 

\begin{Pro}[\cite{rel-dec}]\label{F} Let $X$ and $Y$ be  finite alphabets 
containing at least two letters, then there exists a family $\mathcal{F}$ 
of infinitary rational relations 
which are subsets of $X^\om \times Y^\om$, such that,  for $R \in \mathcal{F}$,  either 
$R = X^\om \times Y^\om$ or $R$ is a  ${\bf \Si^1_1}$-complete subset of $X^\om \times Y^\om$, 
but one cannot decide which case holds. 
\end{Pro}

\begin{Cor}
Let $k$ be an integer $\geq 2$ or $k \in \{\aleph_0^-, \aleph_0\}$. Then it
is undecidable 
to determine whether a given infinitary rational relation  is in the class $RAT_\om(\alpha
\leq k)$ 
(respectively $RAT_\om(\alpha < k)$). 
\nl In particular  one cannot decide whether a given 
infinitary rational relation  is non ambiguous or 
is inherently ambiguous of degree $2^{\aleph_0}$. 
\end{Cor}

\proo Consider the family $\mathcal{F}$ given by Proposition \ref{F} and let  
$R \in \mathcal{F}$. 
\nl  If $R = X^\om \times Y^\om$ then $R$ is obviously non ambiguous but 
 if $R$ is a  ${\bf \Si^1_1}$-complete subset of $X^\om \times Y^\om$ then by 
Theorem \ref{mainthe2} the infinitary rational relation $R$ is inherently ambiguous 
of degree $2^{\aleph_0}$. But one cannot decide which case holds and this ends the proof.   \ep

\hs {\bf  Acknowledgements.}  We thank Dominique Lecomte and Jean-Pierre Ressayre 
for useful discussions and the anonymous referees for useful comments on a preliminary version 
of this paper.

\begin{footnotesize}
 
\end{footnotesize}

\end{document}